\documentclass[iop,apj]{emulateapj} \usepackage{apjfonts} \newif\ifdraft \draftfalse \newif\ifpre \pretrue
\usepackage{graphicx}
\usepackage{xcolor}

\shorttitle{Precise absolute astrometry from VIPS}
\shortauthors{Petrov and Taylor}
\revised{}
\accepted{}
\ifpre
  \submitted{}
  \received{June 13, 2011}
  \revised{July 13, 2011}
  \accepted{July 17, 2011}
\else
  \paperid{}
\fi
\bibpunct{(}{)}{,}{a}{}{,}
\makeindex
\citeindextrue
\newcommand{\n}{\nodata}
\newcommand{\getlength}[1]{\ifx#1\end \let\next=\relax
            \else\advance\count255 by1 \let\next=\getlength\fi \next}
\newcount\switch
\newcommand{\Endmat}{\ifnum\switch=0$\fi}
\newcommand{\ifnularg}[1]{ \count255=0 \getlength#1\end \ifnum\count255=0 }
\newcommand{\ifm}{\makebox{}\ifmmode}

\long\def\ifundefined#1#2#3{\expandafter\ifx\csname
  #1\endcsname\relax#2\else#3\fi}
\newcommand{\beq}   { \begin{eqnarray} }
\newcommand{\eeq}[1]{ \ifnularg{#1} end{eanarray} \else
                      \label{#1}\end{eqnarray}    \fi }
\newcommand{\eeqn}{\nonumber\end{eqnarray}}
\newcommand{\ntab}[2]{ \multicolumn{1}{#1}{#2} }
\newcommand{\nntab}[2]{ \multicolumn{2}{#1}{#2} }
\newcommand{\nnntab}[2]{ \multicolumn{3}{#1}{#2} }

\newcommand{\vex}{\vspace{1ex}}

\newcommand{\PIMA}{$\cal P\hspace{-0.067em}I\hspace{-0.067em}M\hspace{-0.067em}A$ }

\newcommand{\Number}[1]{\ifnum#1<10\relax0\number#1\else\number#1\fi}
\newcommand{\isodate}{
\count151=\time
\divide\count151 by 60
\count151=\count151
\multiply\count151 by 60
\count152=\time
\advance\count152 by -\count151
\divide\count151 by 60
\count152=\count151
\multiply\count151 by 60
\count153=\time
\advance\count153 by -\count151
\Number{\year}.\Number{\month}.\Number{\day}--\Number{\count152}:\Number{\count153}
}
\definecolor{Dred}{rgb}{0.312,0.070,0.070}
\definecolor{Dblue}{rgb}{0.070,0.070,0.312}
\definecolor{Dgreen}{rgb}{0.070,0.312,0.070}
\definecolor{Db}{rgb}    {0.050,0.0,0.320}

\newcommand{\Blb}[1]{\textcolor{Dblue}{\bf #1}}

\ifpre
    \usepackage{hyperref}
    \hypersetup{
            colorlinks=true,
            linkcolor=black,
            filecolor=black,
            citecolor=black,
            urlcolor=blue
           }

     \newcommand{\web}[1]{\Blb{\url{#1}}}
\else
     \newcommand{\web}[1]{#1}
\fi
\newcounter{note}
\let\oldmarginpar\marginpar
\renewcommand\marginpar[1]{\-\oldmarginpar[\raggedleft\footnotesize #1]%
{\raggedright\footnotesize #1}}

\begin{document}

\title{Precise absolute astrometry from the VLBA imaging and polarimetry
       survey at 5 GHz}

\author{L. Petrov}
\affil{ADNET Systems Inc./NASA GSFC, Greenbelt, MD 20771, USA}
\email{Leonid.Petrov@lpetrov.net}
\author{G.B. Taylor}
\affil{Department of Physics and Astronomy, University of New
       Mexico, Albuquerque NM, 87131, USA; Greg Taylor is also an
       Adjunct Astronomer at the National Radio Astronomy Observatory.}
\email{gbtaylor@unm.edu}

\begin{abstract}


  We present accurate positions for 857 sources derived
from the astrometric analysis of 16~eleven-hour experiments from the
Very Long Baseline Array imaging and polarimetry survey at 5 GHz
(VIPS). Among the observed sources, positions of 430 objects were not
previously determined at milliarcsecond-level accuracy. For 95\% of
the sources the uncertainty of their positions ranges from 0.3 to
0.9~mas, with a median value of 0.5~mas. This estimate of accuracy is
substantiated by the comparison of positions of 386 sources that
were previously observed in astrometric programs simultaneously at
2.3/8.6~GHz. Surprisingly, the ionosphere contribution to group
delay was adequately modeled with the use of the total electron
content maps derived from GPS observations and only
marginally affected estimates of source coordinates.


\end{abstract}


\section{Introduction}
\label{s:introduction}

  At high galactic latitudes 70\% (662 of 1043) of the $\gamma$-ray
bright sources detected in the LAT first-year catalogue (1FGL;
\citet{r:abdo10}) are associated with AGN with high confidence ($P >$ 80\%).
The vast majority of the Fermi $\gamma$-ray sources are blazars, with
strong, compact radio emission.  These blazars exhibit flat radio
spectra, rapid variability, compact cores with one-sided parsec-scale
jets, and superluminal motion in the jets \citep{r:mar06,r:tay07,r:lin11}.
In the zone $|b|>10\degr$ radio emission at parsec scales has been detected
with VLBI for 60\% (624 out of 1040) of the 1FGL $\gamma$-ray
sources (Kovalev \& Petrov, 2011 in preparation).

In preparation for the {\it Fermi} mission, an all-sky catalogue
  of compact radio sources, the Combined Radio All-sky Targeted Eight
  gigaHertz Survey (CRATES; \citet{r:hea07}) containing over 11,000
  sources was established. This catalogue is being used with
considerable success in associating {\it Fermi} detections with AGN
\citep{r:abdo10}.  A subset of 1127 sources from CRATES was
  observed with VLBI to form the VLBI Imaging and Polarimetry Survey
  (VIPS; \citet{r:vips}).

The scientific goals for VIPS are:

\begin{enumerate}
   \item To gather
       statistically complete information on large subsamples in order
       to study the apparent properties of objects as a function of the angle
       between the jet and the line-of-sight.

    \par\vspace{-0.5ex}
    \item To explore the underlying patterns in the jets, which can 
       be obscured in images by the peculiarities of individual 
       objects.  In a manner similar to weak lensing, it is possible 
       to average  out these peculiarities.

    \par\vspace{-0.5ex}
    \item To explore the evolution of AGN by studying the size
       (and eventually age) distribution of Compact Symmetric Objects
       (CSOs); (Tremblay et al.(2011), in preparation).

    \par\vspace{-0.5ex}
    \item To search for additional Supermassive Binary Black Hole
       systems (SBBHs).  These can be used to probe the stalling radius
       for black hole aggregation, and to determine the prevalence of mergers
       that might be sources of gravitational radiation.  The same
       observations will also be used to detect or improve the limits
      on milliarcsecond-scale gravitational lenses.
\end{enumerate}

   The original analysis of VIPS observations was made using an automated
approach developed by \citet{r:tay05} based on the software package
Difmap. That pipeline produces images in Stokes parameters $I$, $Q$, and $U$
that were consecutively analyzed for deriving total intensity, polarized 
intensity and electric vector position angle.
These results are discussed in depth in \citet{r:vips}. Originally,
absolute astrometry was not considered as a goal of this project.
Although it was feasible to reduce VLBA data with AIPS and determine
source coordinates in the absolute astrometry mode \citep{r:rdv},
that approach raised the detection limit by a factor of $\sqrt{8}=2.83$,
required intensive manual work, 3--5 working days per experiment, and was not
considered practical for large projects.

   During 2005--2011, significant efforts were made to develop modern
algorithms for processing survey-style experiments and to implement them 
in software. This approach was oriented on unattended astrometric analysis
of an arbitrarily large amount of data and has been described in complete detail in \citet{r:vgaps}. The validation of that approach against 
a set of $500,000$ VLBA observations under regular geodesy and absolute 
astrometry program RDV previously processed with AIPS has proven its 
vitality, high accuracy, lack of systematic errors, and a low detection 
limit approaching the theoretical limit. The amount of time required 
for processing a typical survey experiment was reduced to 2--3 hours, 
making it practical to re-analyze old VLBA experiments and to derive 
positions of observed sources.

In this paper we show the results of our re-analysis of VIPS observations
aimed at absolute astrometry and present the catalogue of source
positions. The motivation of this work was first to add
milliarcsecond-level accurate position information to each data-sheet
associated with the VIPS sample: images at 5~GHz, polarization
measurements, optical photometry from the Sloan Digital Sky Survey,
redshifts, and $\gamma$-ray fluxes. Our second goal was
to extend the VLBA Calibrator list. Since it was known from image
analysis that the majority of VIPS sources are compact, we expect
they will be excellent phase-calibrators as well. Our third goal
was to obtain highly accurate positions of optically bright quasars from
the VIPS list for validation of results from the planned space
astrometry mission \citep{r:gaia}.

In section \ref{s:obs} we briefly describe the observations, sample
definition, and scheduling. In section \ref{s:anal} we outline
the data analysis methodology. The catalogue is presented in
section \ref{s:cat} followed by a brief discussion. Final remarks are
given in the summary.

\section{Observations}
\label{s:obs}

\subsection{Sample Definition}

   To meet the primary goals of VIPS, a relatively large sample
of likely AGN, preferably with data from other wavelength regimes, was
required.  To this end, the Cosmic Lens All-Sky Survey 
\citep[CLASS; ][]{r:mey03} as the parent sample was chosen.  CLASS 
is a VLA survey of $\sim$12,100 flat-spectrum objects ($\alpha > -0.5$ 
between 4.85 GHz and a lower frequency), making it an ideal source of likely
AGN targets to be followed up with the VLBA. The sample was also 
restricted to lie within the survey area, or "footprint" of the Sloan
Digital Sky Survey \citep[SDSS; ][]{r:yor00} as defined at the outset
of VIPS in the fifth data
release of the SDSS \citep[DR5; ][]{r:ade06}.  In DR5 the imaging covers 8,000
square degrees and includes $\sim 2 \times 10^{8}$ objects.
Spectroscopy was obtained as part of the SDSS for $\sim 10^{6}$ of
these objects, about $10^{5}$ of which are quasars. {The source catalog 
was chosen such} that all sources lie on the original SDSS
footprint with an upper declination limit of 65$^{\circ}$ (imposed to
avoid the regions not imaged through DR5).  We
also excluded sources below a declination of 15$^{\circ}$ because it
is difficult to obtain good ($u,v$) coverage with the VLBA for these
objects.  To keep the sample size large but manageable and to obtain a
high detection rate without phase referencing, all CLASS
sources were selected within this area on the sky with flux densities 
at 8.5~GHz greater than 85~mJy, yielding a sample of 1,127 sources.  For
further details see \citet{r:vips}.

\subsection{VIPS Observations}

  The VIPS program was allocated 200 hours of observing time with
the VLBA.  This time was divided into 18 observing sessions of $\sim$11
hours each which ran between January 3 and August 12 2006 (project
BT085). In each session a group of 52--54 target sources were observed
along with the calibrators \object{3C279}, and \object{J1310+3220},
and two other calibrators drawn from the following list:
\object{DA193}, \object{OQ208}, \object{3C273}, \object{J0854+2006},
and \object{J1159+2914}.  In total 958 VIPS sources were observed.
Each VIPS target was observed for approximately 500 seconds divided
into 10 separate scans.  All observations were conducted with four
8~MHz wide, full polarization frequencies centered at 4609, 4679,
4994, and 5095~MHz. These frequencies were chosen in order to allow
for determination of Faraday rotation measures in those sources with
sufficiently polarized components, and in order to improve the
($u,v$) coverage obtained.

  All VLBA observations were scheduled using version 6.05 of
the SCHED program\footnote{See documentation at \\ \hspace{2em}
\web{http://www.aoc.nrao.edu/software/sched}}. Using built-in
data regarding the locations and operation of the VLBA stations,
a special optimization mode ``HAS'' of SCHED automatically produced
a schedule for a list of targets with scan durations, a starting local
sidereal time (LST), and total experiment duration that is optimized both
for ($u,v$) coverage and efficiency.  For each observing run, the
starting LST and scan time per source was varied to produce a schedule
that obtained for the entire duration of 11~hours the vast majority
(if not all) of the required scans for a subset of $\sim\!\!50$ VIPS targets
while minimized slewing time between sources.

  Scan lengths were set to $\sim\!\! 50$ seconds, and the
scheduling software distributed 10 scans of each target source at
approximately equal hour angle intervals. Optimization was done for
the VLBA antenna {\sc pietown} since it is fairly centrally located
to the array. The minimum elevation at that antenna was set to 15 degrees,
and the minimum number of antennas above the horizon was set to 8.  The maximum
deviation from the optimal time for the scan was set at 3 hours.

  Care was also taken to select the correct polarized calibration source(s)
for each run so that it/they would be observed over a wide range
of parallactic angle values while not significantly reducing the
efficiency of the schedule for that run. These sources were also used as
amplitude calibrators for evaluation of antenna complex bandpasses.

\begin{figure}[tbh]
  \centering
  \ifpre
      \includegraphics[width=0.48\textwidth]{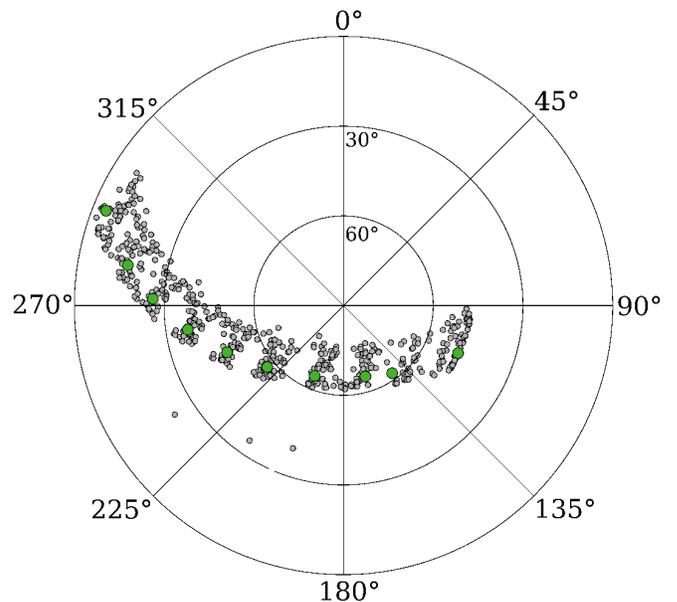}
  \else
      \includegraphics[width=0.48\textwidth]{azel.eps}
  \fi
  \caption{Azimuthal-elevation source distribution over the sky at station
           {\sc hn-vlba} during 11-h long experiment bt085f.
           Source $1609+190$ is highlighted by a big green filled circle.
           Other sources are displayed as small outlined grey circles.}
  \label{f:azel}
\end{figure}

  The schedule optimization that minimized slewing time
resulted in scheduling sources at a relatively wide band across
azimuths and elevations at observing stations. See Figure~\ref{f:azel}
as an example. No elevation cutoff-limit was enforced, and antennas
were used as low as the physical horizon mask permits, down
to $3\degr$.

  Unfortunately, the archive magnetic tape with correlator output from
two experiments, bt085d and bt085e, had deteriorated to the extent that it
could not be read, and the data were completely lost. Therefore, we were
able to re-analyze only 16 out of 18 experiments for a total of 
857~sources.

\section{Data analysis}
\label{s:anal}

%
%
%

  The data were correlated at the Socorro hardware VLBA correlator.
The correlator computed the spectrum of cross correlation and
autocorrelation functions with a frequency resolution of 0.5~MHz and
accumulation intervals 1.96608~s long.

\subsection{Post-correlation analysis}

The procedure of further analysis is described in complete detail in
\citet{r:vgaps}. Here only a brief outline is given. First, the
fringe amplitudes were corrected for the signal distortion in the
sampler and then were calibrated according to measurements of system
temperature and elevation-dependent gain. Phase of the
phase-calibration signal was subtracted from the fringe phases. Then
the group delay, phase delay rate, group delay rate, and fringe phase
were determined for the RR polarization (i.e. right circular
polarization (RCP) at a reference station correlated against the RCP 
at a remote station of a baseline) for all
observations, using the wide-band fringe fitting procedure. These
estimates maximize the sum of the cross-correlation spectrum
coherently averaged over all accumulation periods of a scan and over
all frequency channels in all intermediate frequencies (IFs). 
After the first run of fringe fitting, an observation at each baseline 
with the reference station ({\sc kp-vlba}) with the strongest signal 
to noise ratio (SNR) were used to compute the station-based complex 
bandpass corrections of the RR cross-correlation spectrum. The initial 
bandpass is the cross spectrum coherently averaged over time 
divided by the average of each IF. If the bandpass
characteristics of each IF were constant over an experiment, then
applying the bandpass, i.e. dividing the cross-spectrum by the
bandpass, would make the cross-spectrum averaged over time completely
flat, which reduces the coherence loss. This procedure was repeated
for 12 sources with the strongest SNR at each baseline, and the
results were averaged.

   Next, we computed the polarization complex bandpass. This 
was done for each baseline with a reference station in two steps. First, 
we selected a scan with the highest SNR, applied the initial RR complex 
bandpass and computed group delay $\tau_g$, group delay rate $\dot{\tau_g}$,
phase delay $\tau_p$, and phase delay rate $\dot{\tau_p}$ at the 
reference frequency using RR data. Then we applied the RR complex bandpass to 
the LL data $C^{LL}_{kj}$ at the $k$th accumulation period at the $j$th 
spectral channel and rotated its phase using parameters of the fringe 
fitting to the RR data:
\beq
   \hspace{-1em}
       C^{LL}_{kj} \times e^{-i\,(\, \omega_0
              ( \tau_p \; + \; \dot{\tau}_p (t_k-t_0)) \;\; - \;\;
                2(\psi_1 - \psi_2) \;\; + \;\;
                (\omega_j - \omega_0) \tau_g \,)}.
\eeq{e:e1}
   where $\omega_0$ is the reference frequency, $t_0$ is the fringe
reference time, $\omega_j$ is the frequency of the $j$th spectral
channel, and $t_k$ is time of the $k$th accumulation period.
We also subtracted the doubled difference of parallactic angles
$(\psi_1 - \psi_2)$ at the reference and remote stations
of a baseline. The latter correction accounts for the fact that
the phase rotation due to a change of the receiver orientation with
respect to some reference parallactic angle is opposite for the RCP and LCP
(left circular polarization) signal. The residual amplitudes were
normalized at each IF by its average over all spectral channels.
We approximated the normalized residual complex spectrum with the Legendre
polynomial of the 5th degree. These polynomials define the so-called initial
polarization bandpass. It was found that using polynomials of higher
degree did not provide further improvements.

  Next, we processed 12 observations with the highest SNR at
each baseline with the reference station. We applied the sampler correction,
phase calibration to the RR data, computed parameters of the fringe fitting
to the RR data, applied the sampler correction, phase calibration, performed
the phase rotation according to expression \ref{e:e1} to the LL data of the
same observation, and divided the residual spectrum by the initial complex
polarization bandpass. The phase and the logarithm of the amplitude
of the residual LL spectrum are used as the right-hand side of the observation
equations that approximate the residual complex polarization bandpass
with Legendre polynomials of the 5th degree polynomial for each IF 
and station, except the reference station. This system of linear equations 
is solved using least squares (LSQ). The product of these estimates with 
the initial polarization bandpass forms the final complex polarization 
bandpass.

   The procedure of computing group delays was repeated with the
complex polarization bandpass applied to all the data and using the
Stokes parameter $I$, i.e. the linear combination of RR and LL
cross-spectrum $C_{RR}$ and $C_{LL}$:
$1/2 ( C_{RR} + C_{LL}/P_b )/B_{RR} $, where $B_{RR}$ is the complex
bandpass for the RR data, and $P_b$ is the polarization bandpass.

   If the electronics of every IF were stable during the experiment
and observed sources are completely unpolarized, the SNR of the coherent
sum over time and frequency of the Stokes parameter $I$ would have been
a factor of $\sqrt{2}$ times the SNR when only RR or LL polarization data
are used. Analysis of the SNR ratios of Stokes parameter $I$ data
to the SNR of RR data showed that achieved SNR was on average 96\%
of its theoretical value.

This part of the analysis is done with \PIMA\
software\footnote{Available at \web{http://astrogeo.org/pima}.}.

\subsection{Astrometric analysis}

  The results of fringe fitting the linear combination of RR and
LL data were exported to the VTD/post-solve
VLBI analysis software\footnote{Available at \web{http://astrogeo.org/vtd}.}
for interactive processing of the group delays with SNR high enough to ensure
the probability of false detection is less than 0.001. This SNR threshold
is $5.5$ for the VIPS experiment. A detailed description of the method for
evaluation of the detection threshold can be found in \citet{r:vgaps}.
Next, theoretical path delays were computed according to the state-of-the
art parametric model as well as their partial derivatives. Small
differences between group delays and theoretical path delay were 
estimated using a least squares adjustment to the parametric
model that describes the observations. Coordinates of target sources,
positions of all stations (except the reference station), parameters of the
spline that models residual path delay in the neutral atmosphere in
the zenith direction for all stations, and parameters of another spline
that models the clock function, were solved for in LSQ solutions
that used group delays of the Stokes parameter $I$. Outliers of the
preliminary solution were identified and temporarily barred from
the solution.

   The most common reason for an observation to be marked as an outlier
is a misidentification of the main maximum of the two-dimensional
Fourier-transform of the cross-spectrum. The procedure of fringe
fitting was repeated for observations marked as outliers. Using parameters
of the VLBI model adjusted in the preliminary LSQ solution, we can predict
group delay for outliers with accuracy better than 1~ns. Then we re-run
the fringe fitting procedure for observations marked as outliers in
the narrow search window $\pm 1$~ns within the predicted value of
the group delay. New estimates of group delays with probabilities of 
false detection less than 0.1, which corresponds to the SNR $> 4.6$ for 
the case of a narrow fringe search window, are used in the next step of 
the preliminary analysis procedure. The observations marked as outliers 
in the preliminary solution and detected in the narrow window at the second 
round of the fringe fitting procedure, were tried again. If the new estimate 
of the residual was within 3.5 standard deviations, the observation 
was restored and used in further analysis. Parameter estimation and 
elimination of remaining outliers were then repeated. Finally, the additive 
weight corrections were computed in such a way that these corrections, 
added in quadrature to the a~priori weights computed 
on the basis of group delay uncertainties evaluated by the fringe 
fitting procedure, will make the ratio of the weighted sum of residuals 
to their mathematical expectation close to unity.

  In total, $310\,653$ estimates of group delays out of
$347\,803$ scheduled were used in the solution. The number of observations
of 95\% of the target sources was in the range 200--440, with the median
number 377. Each source was detected, except \object{1058+245}. Detailed
analysis showed that there are two objects in the field of view near
\object{0748+582}. The companion \object{0748+58A} is $72''.8354$ apart.

  The result of the preliminary solution provided the clean dataset of
group delays with updated weights. We used this dataset and all dual-band
S and X data acquired under absolute astrometry and space geodesy 
programs from April 1980 through April 2011, in total 8.2 million 
observations, to obtain final parameter estimation. Thus, the VIPS 
experiments were analyzed by exactly the same way as all other VLBI 
experiments. The estimated parameters were right ascensions and 
declination of all sources, coordinates
and velocities of all stations, coefficients of the B-spline expansion
of non-linear motion for 18 stations, coefficients of harmonic site position
variations of 48 stations at four frequencies: annual, semi-annual, diurnal,
semi-diurnal, and axis offsets for 69 stations. Estimated variables also
included the Earth orientation parameters for each observing session and
parameters of clock function and residual atmosphere path delays in the
zenith direction modeled with the linear B-spline with interval 60 minutes.
All parameters were adjusted in a single LSQ run.

  The system of LSQ equations has an incomplete rank and defines a family
of solutions. In order to pick a specific element from this family, we applied
the no-net rotation constraints on the positions of 212~sources designated
as ``defining'' in the ICRF catalogue \citep{r:icrf98} that required
the positions of these sources in the new catalogue to have no rotation
with respect to the position in the ICRF catalogue. No-net rotation and
no-net translation constraints on site positions and linear velocities
were also applied. The specific choice of identifying constraints was
made to preserve the continuity of the new catalogue with other VLBI
solutions made during last 15 years.

  The global solution sets the orientation of the array with respect to
an ensemble of $\sim\!\!\!5000$ extragalactic remote radio sources.
The orientation is defined by the continuous series of Earth orientation
parameters and parameters of the empirical model of site position
variations over 30 years evaluated together with source coordinates.
Over 400 common sources observed in the VIPS provided a firm connection
between the new catalogue and the old catalogue of compact sources.

\subsection{Accounting for the ionosphere contribution}

   Since the observations were made at a relatively low frequency 
of 5~GHz, one may expect that the dominant source of errors will be the
mismodeled ionosphere contribution. The superior way to alleviate
the frequency-dependent contribution of the ionosphere to group delays
is to record simultaneously at two widely separated frequencies,
i.e. 2.3~GHz (S-band) and 8.6~GHz (X-band) and to use ionosphere-free
linear combinations of these observables in the data analysis. The
residual delay due to high order terms that are not eliminated in the
combination of observables is at a level of several ps \citep{r:iono2nd},
which is negligible.

In the absence of simultaneous dual-band observations, we
computed the ionospheric path delays from the TEC maps derived from
the analysis of Global Positioning System (GPS) observations using the
analysis center CODE \citep{r:scha98}. Using GPS-derived TEC maps for
data reduction of astronomic observations, first suggested by
\citet{r:ros00}, has become a traditional approach for data
processing. However, we should be aware that these path delays are an
approximation of the ionospheric contribution and they cannot account
completely for the contribution of the ionosphere. First, the TEC
model has rather a coarse resolution: 5\degr in spatial coordinates
and $2^h$ in time. Therefore, ionospheric fluctuations on time scales
less than several hours cannot be represented by this model. Second,
the model itself is not precise, since the number of GPS stations that
were used for deriving the model (179 in January 2006) is relatively
small and their distribution around the globe is far from
homogeneous.

  The availability of dual-band observations at the VLBA network during
the period of time overlapping with the VIPS campaign prompted us to use
this dataset for characterizing the residual errors of ionospheric path delays
from GPS TEC maps. Since the ionospheric electron density varies by more
than one order magnitude during a Solar activity cycle, we restricted
our analysis to 15 twenty-four hour experiments under geodetic program
RDV \citep{r:rdv} and VCS6 \citep{r:vcs6} over the period of time from
June 2005 through March 2007. In total, over $130,000$ dual-band
observations were used for comparison. For each observation, we computed
the contribution of the ionosphere to group delay at 4.8~GHz from both
X-band and S-band VLBI group delay observables and from GPS TEC maps. 
The technique of further analysis is described in details in \citet{r:vgaps}. 
Here we briefly outline it.

  For each baseline and each experiment, we computed two statistics:
$RV\!$, which stands for root mean square (rms) of the differences in
the ionospheric group delay between VLBI and GPS measurements, and
$RG = \sqrt{\mbox{rms}_{g1}^2 + \mbox{rms}_{g2}^2}$, where $\mbox{rms}_{gi}$
is the rms of the GPS path delay at the $i$ th station of a baseline during
a session. As we found previously, these statistics are highly correlated.
For each baseline we computed a linear regression between them:
$RV_b = F_b + S_b \cdot RG_b$. The coefficients $F_b$ and $S_b$ depend
on the baseline. Using this regression, we can predict for a given
baseline the rms of the residual contribution of group delay computed
from GPS TEC maps.

\subsection{Error analysis}
\label{s:err}

\begin{figure*}[tbh]
  \centering
  \ifpre
      \includegraphics[width=0.48\textwidth]{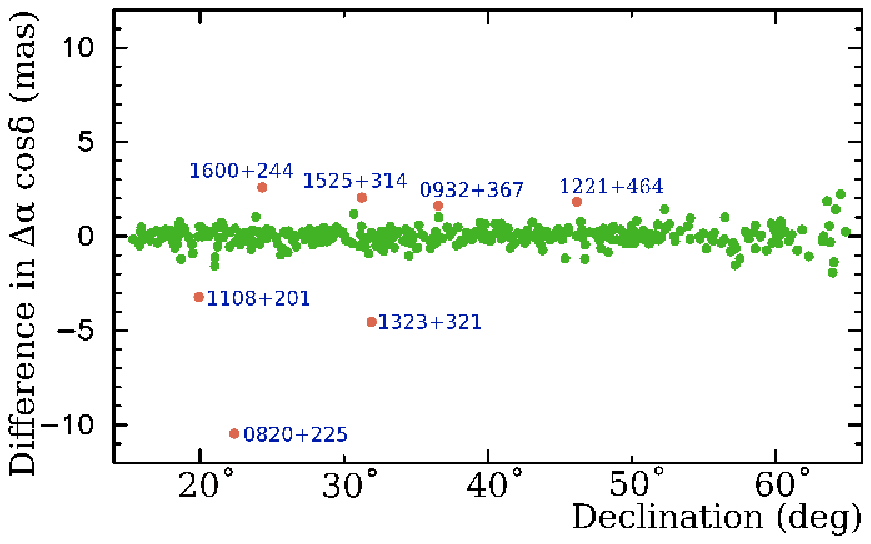}
  \else
      \includegraphics[width=0.48\textwidth]{d_alp.eps}
  \fi
  \hspace{0.02\textwidth}
  \ifpre
      \includegraphics[width=0.48\textwidth]{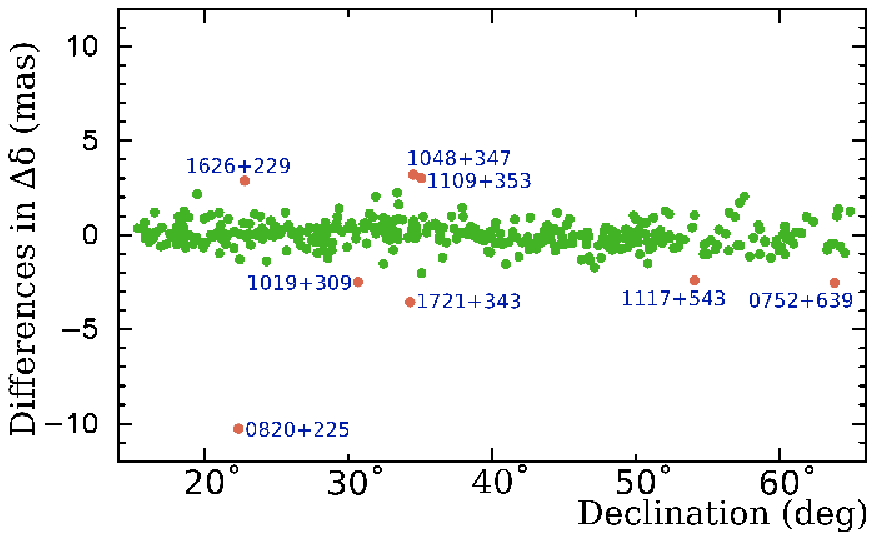}
  \else
      \includegraphics[width=0.48\textwidth]{d_del.eps}
  \fi
  \caption{Differences in positions of common sources derived from
           analysis of VIPS observations at C-band against positions
           derived from analysis of dual-band X/S observations.}
  \label{f:diff}
\end{figure*}

  One half of the sources scheduled for VIPS has been observed 
as part of the VLBA Calibrator Survey 
\citep{r:vcs1,r:vcs2,r:vcs3,r:vcs4,r:vcs5,r:vcs6}
and RDV program \citep{r:rdv} aiming to determination of source positions
in the absolute astrometry mode with the highest accuracy. These observations
were made simultaneously at X and S-bands. We can compare the positions of
common sources from single-band VIPS observations with dual-band observations
and, considering the positions from dual-band observations as the ground
truth, derive the error model.

  We scrutinized the list of 430 common sources and discarded from the list
54 objects with position uncertainties from X/S observations exceeding
1~mas. We are aware that the positions of 386 common sources from the VIPS
catalogue were made from both VIPS observations {\it and} all others
observations. We cannot exclude all common sources from the VIPS solution,
since in that case the matrix of observations will be altered
and properties of such a solution would be significantly different
from the properties of the VIPS catalogue. To circumvent this problem,
we sorted the list of remaining 386 sources in ascending order of right
ascensions and split it into two subsets of 193 sources each, even and odd.
Then we ran two special solutions: solution A that had 193 sources from
the even subset excluded in all experiments, except VIPS, and solution B
that had 193 sources from the odd subset excluded in all experiments,
except VIPS. The solution setup, except the differences in the source list,
was identical to that used for deriving the VIPS catalogue. In all solutions
we estimated positions of 5317 sources. We obtained positions of common sources
from these two solutions derived only from the C-band VIPS observations
and then combined them to form a catalogue of 386 objects.

\begin{figure}[ht]
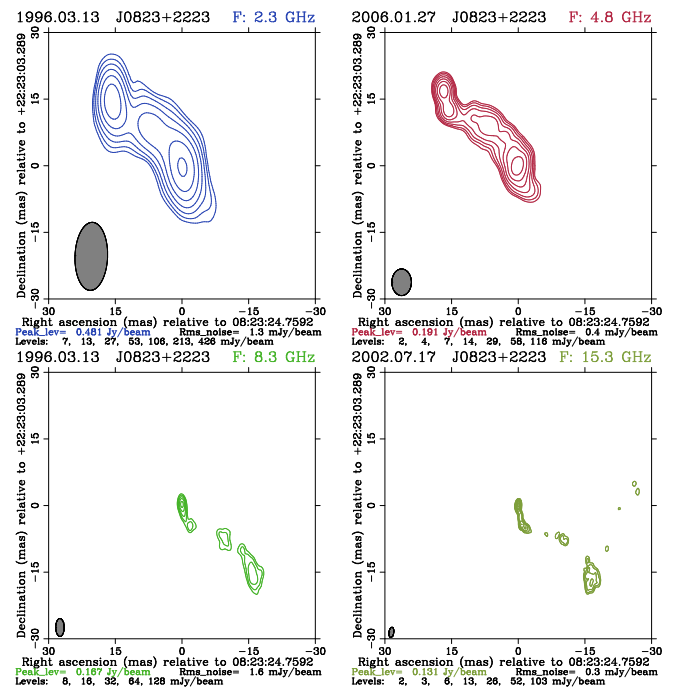

  \centering
      \includegraphics[width=0.23\textwidth]{J0823+2223_S.ps}
      \hspace{0.01\textwidth}
      \includegraphics[width=0.23\textwidth]{J0823+2223_C.ps}
      \hspace{0.01\textwidth}
      \includegraphics[width=0.23\textwidth]{J0823+2223_X.ps}
      \hspace{0.01\textwidth}
      \includegraphics[width=0.23\textwidth]{J0823+2223_U.ps}
  \caption{Image of J0823+2223 (0820+225) in four frequency bands.
           The center of the image is set to the brightest component.
           The images at X and S bands are produced from analysis
           of VCS1 observations (Kovalev, private communication),
           the image at C-band was produced from analysis of VIPS observations,
           and the image at U-band was produced from analysis of observations
           from the MOJAVE program \citep{r:mojave}.
          \label{f:maps}
          }
\end{figure}

   A comparison of these two catalogues is shown in
Figure~\ref{f:diff}. We present the source position differences in
right ascensions and declinations as a function of source declinations.
We see that for a majority of sources, the differences look like a random
noise and do not exhibit any systematic pattern. There are several outliers,
the most notable is \object{0820+225} (See Figure~\ref{f:maps} and
Table~\ref{t:diff}). Analysis of images\footnote{Brightness distributions in
FITS-format as well as images for these and all VIPS sources are available
on the Web from \web{http://www.phys.unm.edu/\~gbtaylor/VIPS/} and from \\
\ifpre
   \web{http://astrogeo.org/vlbi_images}
\else
   \web{http://astrogeo.org/vlbi\_images}
\fi
} of \object{0820+225}, \object{1108+201}, \object{1323+321}, and eight
others revealed a common pattern: all are sources with multiple distinctive
components, and the relative strength of components varies between
bands. Therefore, source positions are related to different parts
of the source. A similar phenomenon was noticed during analysis of differences
of X/S and K-band (24~GHz) positions reported by \citet{r:vgaps}. On average,
1\% sources exhibit this behavior.

\begin{table}
   \caption{The table of sources with the greatest differences
            between VIPS and X/S positions. Quoted flux densities are
            the median correlated flux densities at baselines longer
            than 5000~km.}
   \label{t:diff}
   \begin{tabular}{r l @{} r r l}
     \hline
     Dist  & Err   & B1950-name & J2000-name & Comment \\
     (mas) & (mas) &            &            &         \\
     \hline
     14.6 & 0.5 & 0820$+$225 & J0823$+$2223 & Long, complex jet \\
      5.0 & 0.4 & 1323$+$321 & J1326$+$3154 & CSO \\
      3.6 & 0.5 & 1721$+$343 & J1723$+$3417 & Long, complex jet \\
      3.4 & 0.4 & 1048$+$347 & J1050$+$3430 & Double source (core-jet) \\
      3.4 & 0.4 & 1108$+$201 & J1111$+$1955 & CSO \\
      3.1 & 0.4 & 1109$+$353 & J1112$+$3503 & Weak: 76~mJy at 8~GHz \\
      3.0 & 0.4 & 1600$+$244 & J1602$+$2418 & CSO candidate \\
      2.9 & 0.4 & 1626$+$229 & J1628$+$2247 & Long, complex jet \\
      2.8 & 0.4 & 1019$+$309 & J1022$+$3041 & Long, complex jet \\
      2.6 & 0.5 & 1117$+$543 & J1120$+$5404 & Weak: 38~mJy at 8~GHz \\
      2.6 & 0.7 & 0752$+$639 & J0756$+$6347 & Strong core-jet \\
      2.4 & 0.4 & 0932$+$367 & J0935$+$3633 & Double source (core-jet) \\
      2.2 & 0.5 & 1221$+$464 & J1223$+$4611 & Weak: 67~mJy at 8~GHz \\
      2.1 & 0.4 & 1525$+$314 & J1527$+$3115 & Long, complex jet \\
     \hline
   \end{tabular}
\end{table}

   After removing the outliers, we sought for the ad~hoc variance that,
when added in quadrature with the formal error, makes the ratio of the weighed
sum of the differences to their mathematical expectation close to unity.
We found that the additional variance for right ascensions scaled by cosine
of declination is 0.24~mas and the variance for declinations is 0.31~mas. The
weighted root mean squares (wrms) of the differences in right ascensions
scaled by cosine of declination is 0.44~mas and the wrms of the differences
in declination is 0.58~mas.

   We should note that our estimate of the additive variances provides
an upper limit on the errors. First, de-selecting sources in trial solutions
A and B weakened these solutions with respect to the solution used for
deriving the VIPS catalogue and increased position errors of common sources.
Second, the additive variances 0.24 and 0.31~mas are comparable with errors
of the reference X/S solution. Therefore, the additive variances reflect
contributions of errors of both C-band VIPS and the X/S catalogues.

\section{Source position catalogue}
\label{s:cat}

\tabletypesize{\scriptsize}
\ifpre
  \begin{deluxetable*}{ c l l l r r r r r r r r r l l }
\else
  \begin{deluxetable}{  c l l l r r r r r r r r r l l }
  \rotate
\fi
   \tablecaption{The first 12 rows of the VIPS catalogue of source positions
            of 857 sources. The table columns are explained
            in the text. The full table is available in the electronic
            attachment. \label{t:cat}}
   \tabletypesize{\scriptsize}
   \noindent
   \tablehead{
         & & & & & & & & & & & & & &
         \vex \vex \vex \\
         & & & & & & & & &
         \multicolumn{6}{c}{Correlated flux density (in Jy)}
         \vex \\
         \colhead{Flag }              &
         \nntab{c}{Source Names}      &
         \nntab{c}{J2000 Coordinates} &
         \nnntab{c}{Errors (mas)}     &
         &
         \nntab{c}{S-band} &
         \nntab{c}{C-band} &
         \nntab{c}{X-band}
         \vex \\
         \cline{2-3}
         \cline{4-5}
         \cline{6-8}
         \cline{10-11}
         \cline{12-13}
         \cline{14-15}
         \vex \\
         \colhead{ }   &
         \colhead{IVS  }   &
         \colhead{IAU  }   &
         \colhead{Right $\,$ ascension } &
         \colhead{Declination   }   &
         \colhead{$\Delta \alpha$ } &
         \colhead{$\Delta \delta$ } &
         \colhead{Corr  }  &
         \colhead{\# Obs } &
         \colhead{Total }  &
         \colhead{Unres }  &
         \colhead{Total }  &
         \colhead{Unres }  &
         \colhead{Total }  &
         \colhead{Unres }
         \vex \\
         \ntab{c}{(1)}    &
         \ntab{c}{(2)}    &
         \ntab{c}{(3)}    &
         \ntab{c}{(4)}    &
         \ntab{c}{(5)}    &
         \ntab{c}{(6)}    &
         \ntab{c}{(7)}    &
         \ntab{c}{(8)}    &
         \ntab{r}{(9)}    &
         \ntab{c}{(10)}   &
         \ntab{c}{(11)}   &
         \ntab{c}{(12)}   &
         \ntab{c}{(13)}   &
         \ntab{c}{(14)}   &
         \ntab{c}{(15)}
   }
\startdata
\vex \vex \vex \\
X  & 0552$+$398 & J0555$+$3948 & 05 55 30.805615 & $+$39 48 49.16493 & 0.40 & 0.24 & -0.032 &  1775 &  3.457 &  2.765 & 6.957 &  4.226 &  4.736 &  2.144  \vex \\
C  & 0716$+$450 & J0719$+$4459 & 07 19 55.511712 & $+$44 59 06.84827 & 1.68 & 1.52 & -0.059 &    98 &  \n    &  \n    & 0.151 & <0.02  &  \n    &  \n     \vex \\
C  & 0722$+$393 & J0726$+$3912 & 07 26 04.737260 & $+$39 12 23.32117 & 0.55 & 0.57 & -0.300 &   302 &  \n    &  \n    & 0.076 &  0.045 &  \n    &  \n     \vex \\
C  & 0722$+$415 & J0726$+$4124 & 07 26 22.420172 & $+$41 24 43.66844 & 0.53 & 0.43 & -0.217 &   362 &  \n    &  \n    & 0.106 &  0.090 &  \n    &  \n     \vex \\
C  & 0722$+$615 & J0726$+$6125 & 07 26 51.681732 & $+$61 25 13.68060 & 0.90 & 0.42 & -0.017 &   336 &  \n    &  \n    & 0.096 &  0.088 &  \n    &  \n     \vex \\
X  & 0723$+$488 & J0727$+$4844 & 07 27 03.100595 & $+$48 44 10.12669 & 0.52 & 0.34 & -0.004 &   362 &  0.447 &  0.357 & 0.263 &  0.222 &  0.330 &  0.280  \vex \\
X  & 0729$+$562 & J0733$+$5605 & 07 33 28.615274 & $+$56 05 41.73927 & 2.34 & 1.98 & -0.238 &    86 &  0.099 &  1.269 & 0.113 &  0.022 &  \n    &  \n     \vex \\
C  & 0731$+$595 & J0735$+$5925 & 07 35 56.300840 & $+$59 25 22.11156 & 0.94 & 0.48 &  0.022 &   322 &  \n    &  \n    & 0.053 &  0.037 &  \n    &  \n     \vex \\
X  & 0733$+$300 & J0736$+$2954 & 07 36 13.661088 & $+$29 54 22.18498 & 0.41 & 0.42 & -0.038 &   344 &  0.464 &  0.409 & 0.273 &  0.158 &  0.302 &  0.178  \vex \\
C  & 0733$+$287 & J0736$+$2840 & 07 36 31.198545 & $+$28 40 36.80515 & 2.05 & 1.97 & -0.144 &    58 &  \n    &  \n    & 0.039 &  0.025 &  \n    &  \n     \vex \\
X  & 0733$+$261 & J0736$+$2604 & 07 36 58.073689 & $+$26 04 49.94573 & 0.41 & 0.42 & -0.034 &   351 &  0.336 &  0.086 & 0.240 &  0.108 &  0.326 &  0.215  \vex \\
C  & 0734$+$269 & J0737$+$2651 & 07 37 54.975269 & $+$26 51 47.44790 & 0.54 & 0.60 & -0.141 &   302 &  \n    &  \n    & 0.071 &  0.051 &  \n    &  \n     \vex \\
\vspace{1.5ex}
\enddata
\tablecomments{Table~\ref{t:cat} is presented in its entirety in the electronic
               edition of the Astronomical Journal. A portion is shown here
               for guidance regarding its form and contents. Units of right
               ascension are hours, minutes and seconds. Units of declination
               are degrees, minutes and seconds.
}
\ifpre
  \end{deluxetable*}
\else
  \end{deluxetable}
\fi

  Table~\ref{t:cat} displays 12 out of 857 rows of the VIPS
catalogue of source positions. The full table is available in the
electronic attachment. Column 1 contains the source flag: ``X'' if
the sources was previously observed at X and S-bands under absolute
astrometry programs, ``C'' otherwise. Columns 2 and 3 contain
the J2000 and B1950 IAU names; column 4, and 5 contain right ascensions 
and declinations. Columns 6 and 7 contain positional uncertainties
in right ascension (without multiplier $\cos \delta$) and in declinations
with noise added in quadrature to their formal errors. Column 8 contains
correlation coefficients between right ascension and declination,
column 9 contains the total number of observations used in position
data analysis. Columns 10 through 15 contain the median estimates
of the correlated flux density over all experiments of a given source in two
ranges of baseline projection lengths: less than 900 km, and longer
than 5000~km. The first estimate is close to the total correlated
flux density, the second estimate is close to the correlated flux
density of the unresolved component. The estimates of the median
correlated flux densities are given for three bands: S-band
(columns 10,11), C-band (columns 12,13), and X-band (columns 14,15).
The estimate of the flux density at C-band was computed from analysis
of VIPS data by \citet{r:vips}, estimates at S and X band are from
re-analysis of VCS and RDV observations. If there were no detections
at the range of the baseline projection lengths, then $-1.000$ is put
in the table cell.

\begin{figure}[tbh]
  \centering
  \ifpre
      \includegraphics[width=0.48\textwidth]{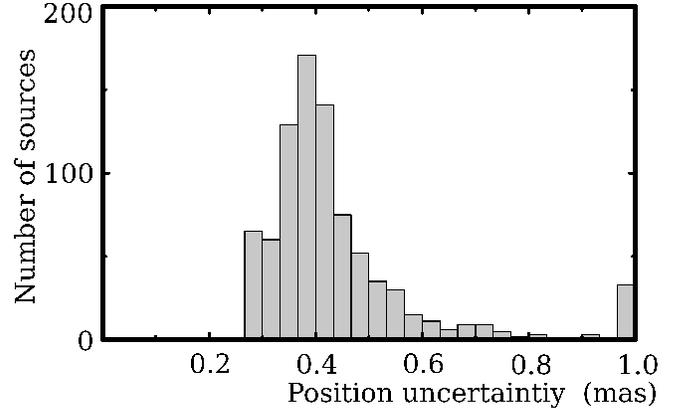}
  \else
      \includegraphics[width=0.48\textwidth]{pos_hist.eps}
  \fi
  \caption{The histogram of semi-major axes of inflated position uncertainties
           of the VIPS source position catalogue. The last bin shows
           uncertainties exceeding 1~mas.}
  \label{f:hist}
\end{figure}

  The histogram of the semi-major axes of position uncertainties is
given in Figure~\ref{f:hist}.

\section{Discussion}
\label{s:disc}

  The astrometric accuracy of VIPS exceeded our expectations. Although 
absolute astrometry was not considered at all during design of the experiment,
position uncertainties of target sources observed in single band VIPS
experiments at 5~GHz turned out better than the position uncertainties
of dedicated VCS experiments. The contribution of the residual ionosphere
path delay after applying the GPS-derived TEC model was small and did
not affect the positional accuracy significantly.

  Several factors contributed to this success. First, the wide-band
frequency setup spanned over 494~MHz was used. This is essential.
If a contiguous bandwidth of 32~MHz had been recorded, as is often
selected in imaging experiments, group delays would have been a factor
of 22 less precise, and as a result, the source position accuracy would
have been more than one order of magnitude worse. Second, each source was
scheduled in 10 scans equally distributed over the hour angle of the array
center (station {\sc pietown}). The schedule was optimized to get good
($u,v$)-coverage with no elevation cutoff enforced. Observations at elevations
as low as $3\degr$ were scheduled. Good ($u,v$)-coverage also provides 
a good separation among variables of interest such as source coordinates, 
as well as among nuisance parameters such as the clock function. Frequent
scheduling of low-elevation observations improves the reliability of estimates
of residual atmosphere path delay in zenith parameters, since it helps
to decorrelate these parameters with station positions and clock function
parameters. Third, the integration time was correctly predicted, so the vast
majority of sources were detected at both short and long baselines.
A lack of observations at long baselines reduces the accuracy of source
positions significantly. Finally, the experiments were carried out during
the Solar minimum when ionosphere disturbances were exceptionally low.

  Of the 14 sources with significant discrepancies between their
position measured in VIPS and that derived from dual-band X/S
observations (see Table~\ref{t:diff}, the majority (10/14) have complex
structure in which the brightest component is a function of angular
resolution and/or observing frequency. This includes three Compact
Symmetric Objects which generally have more complex structure than
core-jets. Of the remainder, three sources are core-jets,
but quite weak ($<$ 100 mJy/beam), so it is likely that in this case the
errors are less systematic and more the result of low SNR.
The single remaining source, VIPS \object{J0756+6347}, is a strong and
ordinary core-jet and the reason for the discrepancy in positions is
unclear.

\section{Conclusions}
\label{s:concl}

  We have determined positions of 857 sources observed in 16 VIPS
experiments. Milliarcsecond-accurate coordinates of 430 objects were
determined with VLBI for the first time.

  We found that during the Solar minimum the residual ionosphere
contribution to the group delay after applying the reduction for
path delay from GPS TEC map is small enough not to cause systematic
errors exceeding 0.3~mas at frequencies as low as 5~GHz.

  Our error analysis showed that for 95\% of the sources the uncertainty
of their positions range from 0.3 to 0.9~mas. This estimate is substantiated
by comparison with a large set of 386 VIPS sources that were also observed
in X/S absolute astrometry programs. The position uncertainty of the VIPS
catalogue is better than the uncertainty of dedicated VLBI Calibration
Surveys.

\acknowledgments

We thank an anonymous referee for constructive suggestions.
The National Radio Astronomy Observatory is a facility
of the National Science Foundation operated under cooperative
agreement by Associated Universities, Inc.

{\it Facilities:} \facility{VLBA (project code BT085)}.

\end{document}